\documentclass[12pt,letterpaper]{article}
\usepackage[]{graphics,epsfig}
\newbox\grsign \setbox\grsign=\hbox{$>$} \newdimen\grdimen \grdimen=\ht\grsign
\newbox\simlessbox \newbox\simgreatbox
\setbox\simgreatbox=\hbox{\raise.5ex\hbox{$>$}\llap
     {\lower.5ex\hbox{$\sim$}}}\ht1=\grdimen\dp1=0pt
\setbox\simlessbox=\hbox{\raise.5ex\hbox{$<$}\llap
     {\lower.5ex\hbox{$\sim$}}}\ht2=\grdimen\dp2=0pt
\def	\gtsim{\mathrel{\copy\simgreatbox}}	%gt.approx.eq.to
\def	\ltsim{\mathrel{\copy\simlessbox}}        %lt.approx.eq.to

\def	\kms	{\,{\rm km\,s}^{-1}}
\def	\micron	{\,\mu{\rm m}}
\def	\nm	{\,{\rm nm}}
\begin{document}
\twocolumn
%\vspace*{-15em}
\centerline{to appear in {\it Encyclopedia of Astronomy \& }}
\centerline{{\it Astrophysics} (IOP and MacMillan)}
%\vspace*{14em}
\section*{\centerline{Interstellar Grains}}
\vspace*{-1em}
\centerline{B. T. Draine, Princeton University}
\vspace*{0.5em}
\subsection*{Overview}
Submicron solid particles are dispersed through interstellar gas.
These {\it interstellar grains} absorb and scatter light, thus 
shielding some regions from ultraviolet radiation, but also limiting
our ability to detect photons which have been emitted by
astronomical objects.
Dust grains reradiate absorbed energy in the infrared, thus contributing
to the overall emission spectrum of astronomical systems ranging from
dusty disks around stars to ultraluminous starburst galaxies.

A naked-eye view of the sky from a dark site on a clear summer night reveals
dramatic dark patches in the Milky Way.  These dark regions are not due
to a deficiency of stars -- they are instead the result of obscuration
by dust clouds interposed between the Earth and distant
stars.
The obscuration tends to be greater at shorter wavelengths; as a result,
the light reaching us from distant, obscured stars is ``reddened''.
This reddening by interstellar dust can be understood as arising from
scattering and absorption by a population of interstellar submicron
dust grains.

The grain population spans a range of sizes, 
from molecules\footnote{
	Since all grains are ``molecules'', it is natural to consider
	small molecules as the small-size end of the overall grain
	population.
	}
containing only tens of atoms to
particles as large as $\sim0.3\micron$, containing $\sim10^{10}$ atoms.
Most of the grain mass appears to be 
due to two types of solid, in approximately
equal amounts: (1) amorphous silicate mineral, and (2)
carbonaceous material.
A number of elements -- including silicon and iron -- are primarily
in solid form in the interstellar medium.
Approximately 2/3 of the interstellar carbon in diffuse clouds is
in solid form (see {\scriptsize INTERSTELLAR ABSORPTION LINES}).

It is of course important to characterize the wavelength-dependent
interstellar extinction so that astronomical observations can
be ``corrected'' for the obscuring effects of dust.
In addition, the infrared emission from dust grains
provides a valuable probe of dense regions, and the dust grains
themselves play important roles in interstellar chemistry (shielding
from ultraviolet radiation, and catalyzing the formation of H$_2$),
interstellar gas dynamics (radiation pressure forces on dust grains,
and coupling of charged dust grains to magnetic fields)
and heating and cooling of interstellar gas.
Dust grains are central to many 
problems in modern astrophysics.

\subsection*{Observational Evidence: Summary}

There are many different astronomical phenomena which both reveal the
existence of interstellar dust grains, and provide information allowing
us 
to infer the properties of this dust.
Some of the information is quite direct, due to absorption, scattering,
or emission of light by the grains:
\begin{itemize}
\item Wavelength-dependent extinction -- attenuation and ``reddening'' of 
the light from distant stars due to
intervening dust
(see Figure \ref{fig:extcurvs}).
\begin{figure}[h!]
\epsfig{file=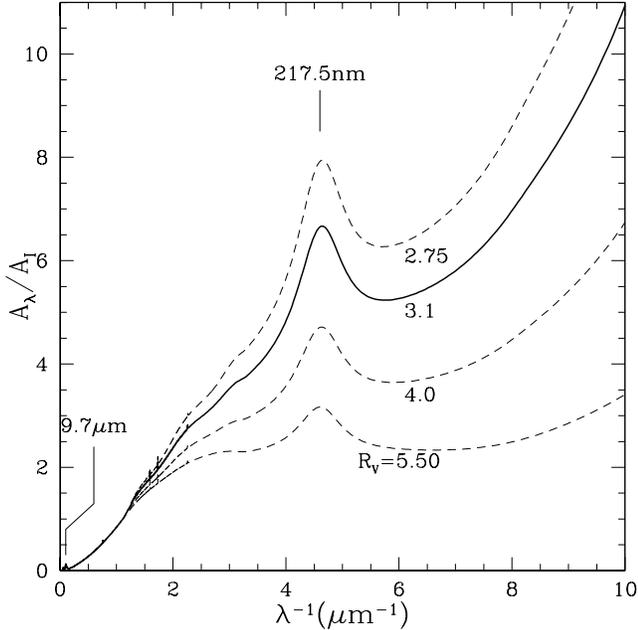,width=\hsize,height=\hsize}
\caption{\small Wavelength-dependent extinction, normalized to the extinction at
$I=900 \nm$, for different types of clouds, identified by the value of
$R_V\equiv A_V/(A_B-A_V)$, where $V=550 \nm$ and $B=440 \nm$.
The average extinction for diffuse clouds is characterized by
$R_V\approx 3.1$.  Dense gas near the surfaces of molecular clouds
can have $R_V$ as large as 5.5.
The extinction at $I$ is approximately proportional to $N_{\rm H} =
N({\rm H}) + 2N({\rm H}_2) + N({\rm H}^+)$, with
$A_I/N_{\rm H}\approx 2.6\times10^{-26}{\rm\, m}^2/{\rm H}$.
\label{fig:extcurvs}
}
\end{figure}
\item Spectroscopic features in the extinction.
There are a number of extinction features, 
including:
\begin{itemize}
\item A strong and very broad extinction ``bump'' at 217.5nm
(see Figure \ref{fig:extcurvs}),
probably due to carbonaceous material, perhaps graphite.
\item Infrared
extinction features at 9.7$\mu$m and 18$\mu$m, almost certainly due
to silicates.
\item A number
of weaker ``diffuse interstellar bands'' (see Figure \ref{fig:dibs}), 
the strongest of which are
at $443\nm$ and $578\nm$, and which remain generally unidentified.
\begin{figure}[h!]
\epsfig{file=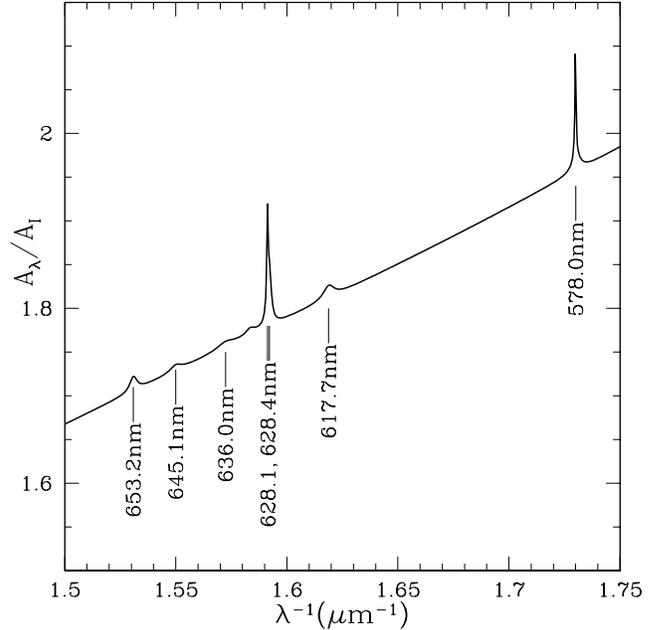,width=\hsize,height=\hsize}
\caption{\small 
	A portion of the extinction curve showing some of the
	``diffuse interstellar band'' extinction features, labelled
	with their respective wavelengths.
	These diffuse bands may 
	be due to impurities in grains, or to ``free-flying''
	large molecules/ultrasmall grains.
	\label{fig:dibs}
	}
\end{figure}
\item An absorption feature at 3.4$\micron$, seen in diffuse clouds,
presumably due to the C-H stretching mode in aliphatic hydrocarbons.
\item A number of absorption 
features, seen only in molecular clouds,
due to ice mantles which apparently coat the grains in these regions.
The strongest such feature is a 3.1$\mu$m feature attributed to
H$_2$O ice.
\end{itemize}
\item Polarization of starlight -- preferential attenuation of one
linear polarization over another by aligned interstellar dust grains,
so that initially unpolarized light from a star is partially
polarized by the time it reaches the Earth
(see {\scriptsize POLARIZATION OF STARLIGHT}).
\item Reflection nebulae -- dust clouds which are relatively close to
bright stars, so that the starlight reflected by dust grains near
the cloud surface renders the cloud visible
(see {\scriptsize REFLECTION NEBULAE}).
\item X-ray haloes around X-ray point sources located behind interstellar
dust clouds.   The haloes result from small-angle scattering of X-rays
by interstellar dust grains.
\item Infrared emission from dust grains heated by interstellar
starlight (see Figure \ref{fig:iremission}).
\begin{figure}[h!]
\epsfig{file=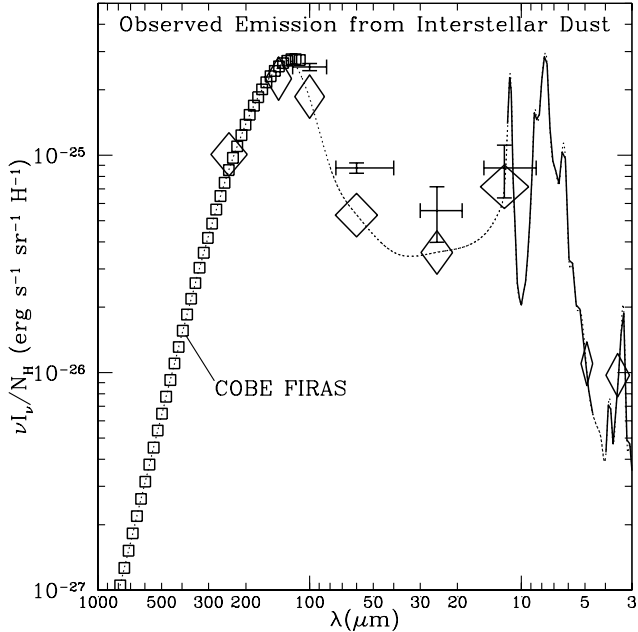,width=\hsize,height=\hsize}
\caption{\small Infrared emission from interstellar dust grains from
dust in diffuse clouds, per
H nucleon.
Crosses indicate data from the {\it InfraRed Astronomy Satellite} ({\it IRAS}) 
at 100, 60, 25, and 12$\micron$.
Squares indicate data from the FIRAS instrument on the
{\it COsmic Background Explorer (COBE)}.
Diamonds indicate data from the {\it DIRBE} instrument on {\it COBE}.
The heavy solid line at 12 -- 5$\micron$ and
4.6 -- 3$\micron$ is the spectrum measured by the {\it InfraRed Telescope
in Space (IRTS)}.
\label{fig:iremission}
}
\end{figure}
\item Infrared emission features indicative of aromatic
hydrocarbons.
The strongest features from the general interstellar medium 
are at 11.3$\mu$m, and 7.7$\mu$m (see Figure \ref{fig:iremission}).
\item Emission features at 18$\mu$m and 9.7$\mu$m, generally thought to
be due to silicates, from dust in regions
with radiation fields $\gtsim 10^4$ times stronger than
the average starlight background.
\item There is strong evidence for emission from interstellar grains
in the far-red, presumably fluorescence following absorption of a shorter
wavelength photon.
\item In the local interstellar medium, the dust-to-gas ratio appears
	to be approximately constant -- the dust follows the gas.
\end{itemize}
Although limited, we also have some quite direct evidence concerning
interstellar grains:
\begin{itemize}
\item Relic interstellar grains found in primitive ``carbonaceous
chondrite'' meteorites.  
\item Impacts with interplanetary probes 
of interstellar grains passing through the
solar system.
\end{itemize}
In addition, there are a number of astrophysical 
phenomena which provide indirect
information about interstellar grains:
\begin{itemize}
\item Underabundances, relative to solar abundances, of certain
elements in interstellar gas -- 
including iron, silicon, magnesium, and carbon
(see {\scriptsize INTERSTELLAR ABSORPTION LINES}).
These elements are presumed to be underabundant in the gas phase because
a large fraction of the atoms are locked up in interstellar grains.
\item The presence in different parts of the
interstellar medium of molecular hydrogen (H$_2$) 
with abundances far exceeding what could be produced
by purely gas-phase processes.
The required rate of  
of H$_2$ catalysis on interstellar grain surfaces provides information
on interstellar grain properties, including total surface area.
\end{itemize}

\subsection*{Extinction by Interstellar Dust Grains}

The ``extinction curves'' shown in Figure \ref{fig:extcurvs} represent
the wavelength-dependent extinction $A_\lambda$ at wavelength $\lambda$
observed on different kinds of
sightlines.
Careful studies of the extinction on many lines of sight
reveals that the extinction curves can be approximately described by
a one parameter family of extinction curves, and the curves shown in
Figure \ref{fig:extcurvs} are obtained from fitting functions originally
developed by Cardelli, Clayton, \& Mathis (1989) and others.
It is convenient to take $R_V\equiv A_V/(A_B-A_V)$
as the parameter, where $V=550\nm$ and $B=440\nm$.
Values of $R_V$ as small as 2.75, and as large as 5.5, are observed
in different regions.

Grains in diffuse regions tend to have $R_V\approx3.1$, while larger
values of $R_V$ tend to be seen when studying the extinction by
dust in dense clouds.  Larger values of $R_V$ -- corresponding to
``greyer'' extinction -- are indicative of
larger grains.
The observed tendency for larger values of $R_V$ to be found in
denser regions strongly suggests that characteristic grain sizes are
larger in these regions.
It can be shown that the increased values of $R_V$ cannot be explained
simply by accretion of atoms and molecules from the gas phase -- because
of the very large amount of surface area contributed by the smaller grains,
accretion from the gas phase would result in only a very small increase
$\Delta a$ in the radii of all grains, with minimal effect on the extinction
at $B$ and $V$.
Instead, small grains must coagulate with
larger grains to convert diffuse cloud dust to dense cloud dust.

\subsection*{Extinction Features}
The strongest spectroscopic feature in the extinction curve is
a conspicuous ``bump'' at $217.5$nm (see Figure \ref{fig:extcurvs}). 
This feature has approximately the wavelength
and width expected for small (radius $a\ltsim 15$nm) 
particles of graphite.
While identification of the feature is still not
certain, it appears highly likely that it is due to small carbonaceous
particles with $sp^2$ carbon-carbon bonds as in
graphite and polycyclic aromatic hydrocarbons.
The strength of the 217.5nm feature requires that $\sim$15\% of the
solar abundance of carbon be present in small ($a\ltsim15$nm) particles.

Between 1.32$\micron$ and 400 nm there are many weak diffuse extinction
features, known as the ``diffuse interstellar bands'', or DIBs.
Approximately 300 such features have been identified, with
full width at half-maximum
(FWHM) ranging
from $\sim$0.05nm up to $\sim$4nm.
The strongest DIB is at 442.9nm.

The first DIBs were recognized in 1934 by Merrill, yet they remain
essentially unidentified to this date.
They must be due to either ``impurities'' within dust grains, or
to small free-flying molecules (i.e., ultrasmall dust grains).
A few of the weaker features have been identified recently 
as electronic transitions of C$_7^-$, and it seems likely that many other
of the DIBs are also produced by small hydrocarbon molecules, either neutral
or charged (either positively or negatively).

A broad extinction feature at 3.4$\micron$ is attributed to the C--H
stretching mode in aliphatic (chainlike) hydrocarbons.
This feature is present in the diffuse interstellar medium.

There are strong infrared extinction features peaking at
9.7$\micron$ and 18$\micron$ which are almost certainly due to amorphous
silicates with a composition approximating that of olivine
(Mg$_x$Fe$_{2-x}$SiO$_4$).
In regions where the dust is hot (e.g., circumstellar dust, or the dust near
the Trapezium in the Orion Nebula), these features appear in emission.
In some circumstellar dust shells and disks (e.g., the dusty
disk around the Herbig Ae/Be star HD 100546), sharp features
characteristic of crystalline silicates appear in emission, but these
features have not been detected in either emission or absorption in the
interstellar medium, indicating that the bulk of interstellar silicates
are amorphous.

In dark clouds, a number of additional features appear in the infrared
extinction, presumably due to growth of molecular ice mantles on
the refractory dust grain cores.
The strongest feature is at 3.08$\micron$ and is due to amorphous
H$_2$O ice.
Additional features have been identified as frozen CH$_3$OH (3.53$\micron$),
CO (4.67$\micron$),
CH$_4$ (7.65$\micron$),
and
CO$_2$ (15.2$\micron$).

\subsection*{Alignment of Dust Grains}

When initially unpolarized 
starlight passes through the dusty interstellar medium, it acquires
both linear and circular polarization.
The linear polarization is due to preferential attenuation of one
linear polarization over the other, due to a population of nonspherical
dust grains which are somehow aligned.
The linear polarization peaks near $V=550$nm, with peak polarization
$P\approx0.03A_V$.
The circular polarization, generally quite small, 
is due to conversion of linear
to circular polarization when the alignment direction of the dust grains
undergoes a twist along the line-of-sight to the source.

Spinning dust grains have magnetic moments antiparallel to their angular
velocities, and the angular momentum therefore precesses around the
local galactic magnetic field.
As a result, the observed direction of starlight linear polarization must be
either parallel or perpendicular to the projection of the magnetic field
on the sky.  Theories of grain alignment lead us to expect the grain
angular momentum to tend to align with the magnetic field, and the ``long''
axis of the grains to tend to be perpendicular to the angular momentum.
The long axis of the grain therefore tends to be perpendicular to
the magnetic field, and as a result light becomes linearly polarized 
{\it parallel} to the projection of the magnetic field on the plane of
the sky.

Our understanding of the physics of dust grain alignment is not yet
complete, but it appears that the observed alignment in diffuse clouds is
produced mainly by
a combination of paramagnetic dissipation (originally
proposed by Davis and Greenstein in 1951) and radiative 
torques on irregular
dust grains due to anisotropic starlight.
In some cases the alignment may result from drift of the dust grains
through the gas cloud.

\subsection*{Scattering of Light by Dust Grains}

Reflection nebulae such as NGC2023 or the beautiful filamentary structures
near the Pleaides show that dust grains
scatter starlight.
While most conspicuous when a bright star is located near the surface
of a dense clouds, reflected starlight also manifests itself as
the ``diffuse galactic light'' -- reflected starlight seen in all
directions in the sky where there is dust.
Measurements of the surface brightness of reflection nebulae, or of
the diffuse galactic light, provide constraints on the scattering properties
of interstellar dust grains -- both the total ``albedo'' 
$\omega = C_{sca}/C_{ext}$, 
and the angular dependence of the scattering, often
characterized by $g\equiv\langle\cos\theta\rangle$, where
$\theta$ is the scattering angle.
At visible wavelengths, the albedo $\omega\approx0.5$ and the
grains are moderately forward-scattering, $g\approx0.5$, consistent with
current models for interstellar grains which reproduce the wavelength-dependent
extinction as well as the infrared emission from dust.

Scattering by dust grains can also be observed at X-ray wavelengths.
A grain is essentially transparent to $h\nu\gtsim 0.5$keV X-ray photons.
Since the refractive index of the grain material is very close to 1, the
scattering can be calculated in the ``Rayleigh-Gans'' approximation, and
one finds that only small-angle scattering is expected.
X-ray halos around compact X-ray sources have been imaged by
the {\it Einstein} and {\it ROSAT} observatories; the observed
X-ray halos appear to be approximately consistent with the scattering
expected for a grain model developed to account for the optical-UV
extinction curve.

\subsection*{Dust Grain Luminescence}

Observations of reflection nebulae (e.g., NGC 7023)
as well as the general interstellar diffuse clouds or
``cirrus'' appear to show evidence of far-red continuum radiation in excess of
what is expected from simple scattering of starlight by dust grains.
The far-red emission peaks near $\sim$700nm, and has a spectrum resembling
the luminescence from hydrogenated amorphous carbon illuminated by
$\lambda \ltsim 550$nm radiation.
This suggests that some of the grain material may resemble
hydrogenated amorphous carbon, although ultrasmall silicon grains have
also been proposed as the source of the emission.
Even if it is assumed that hydrogenated amorphous carbon is responsible for
the observed emission, an accurate estimate of the quantity of material
required is not possible, since luminescence efficiencies of hydrogenated 
amorphous carbon in the laboratory depend
on the preparation of the sample.

\subsection*{Infrared Emission from Dust Grains}

Starlight is in part absorbed by dust grains, and the absorbed energy is
reradiated by dust grains in the infrared.
The observed emission spectrum for interstellar dust is shown in
Figure \ref{fig:iremission}.
It consists of far-infrared emission peaking at $\lambda\approx 140\micron$,
plus substantial emission at shorter wavelengths.

The emission at
$\lambda < 12\micron$ shows conspicuous emission features at 11.3,
8.6, 7.7, 6.2, and 3.3$\micron$; emission peaks at these same wavelengths
have also been observed from reflection nebulae, planetary nebulae,
HII regions, and circumstellar dust.
The emission features have
been identified as characteristic of polycyclic aromatic hydrocarbons
(PAHs): C-H stretch (3.3$\micron$), C-C stretch (6.2 and 7.7$\micron$),
in-plane C-H bend (8.6$\micron$) and out-of-plane C-H bend (11.3$\micron$).
Variations in the relative strengths and precise wavelengths 
of these features from one object
to another may be due to changes in the PAH mixture, including changes
in the fraction which are positively or negatively charged.

The ``classical'' grains with radii $0.01 - 0.3\micron$ radiate a
spectrum characteristic of thermal emission at the (steady) temperature
of the grain, $\sim$15--20~K for grains in the diffuse interstellar medium.
This accounts for nearly all of the emission at $\lambda > 60\micron$.
For very small grains, however, absorption of a single starlight photon
can appreciably change the grain temperature.
For example, a single photon of energy $h\nu=10$eV can heat a 
230 atom graphite grain to a peak temperature $T\approx 300$K.
At this temperature the grain can radiate effectively at wavelengths
as short as $\lambda\approx8\micron$.
The observed emission feature near $7.7\micron$ could therefore be due
to thermal emission from grains with $\sim 100 - 300$ atoms.
The relatively large amount of power radiated at $\lambda < 10\micron$
requires that the grains with $\ltsim 300$ atoms account for an appreciable
fraction ($\gtsim 15$\%) of the total absorption of starlight by interstellar
grains.

\subsection*{Microwave Emission from Dust Grains}

Sensitive observations of the cosmic microwave background radiation have
revealed 10--60 GHz emission from interstellar matter with
intensities greatly exceeding what
would be expected from an extrapolation of the thermal far-infrared
emission to these lower frequencies.
It appears likely that the observed 10--60 GHz emission is largely 
rotational electric dipole emission from very rapidly rotating ultrasmall
grains, although a fraction of the radiation could be thermal emission
from grains containing materials that are ferromagnetic
(e.g., metallic Fe inclusions) or ferrimagnetic (e.g., magnetite
Fe$_3$O$_4$).

\subsection*{Interstellar Dust Grains in the Solar System}

The solar system was formed out of interstellar gas and dust
approximately 4.5 billion years ago.
The formation of planetesimals and planets was accompanied in many
cases by high temperatures and violent conditions, and most interstellar
dust particles were destroyed.  However, the class of meteorites known
as carbonaceous chondrites (see {\scriptsize METEORITES}) contain small
particles with unusual isotopic ratios (see {\scriptsize ISOTOPIC ANOMALIES})
which indicate that they did not
form in the solar nebula, but rather must have been formed in a region with
an anomalous composition (e.g., outflow from an evolved star) long
before the formation of the solar system.
Therefore these particles must have been part of the interstellar grain
population prior to the formation of the solar nebula.

To date, 5 different type of presolar grains have been isolated and identified
(see Table \ref{tab:meteoritic}).
The evidence that the grains listed in Table \ref{tab:meteoritic}
were truly interstellar is compelling, but it is important to realize that
they apparently do not include typical interstellar grains, for the
simple reason that the
procedures used to isolate interstellar grains in meteorites are
designed to deliberately
destroy silicate material (which comprises the bulk of the carbonaceous
chondrite meteorite ``matrix'').
These laboratory procedures are therefore {\it not} going to find interstellar
silicate grains even if they are present.

\begin{table}[h!]
\caption{\bf Interstellar Grains in Meteorites}\label{tab:meteoritic}
\begin{tabular}
{l c c}
\hline
Composition&diameter($\micron$)&Abund.\footnotemark[2]\\
\hline
C (diamond)&			0.002&		$5\times10^{-4}$\\
SiC\footnotemark[3]&		0.3-20&		$6\times10^{-6}$\\
C (graphite)\footnotemark[3]&	1-20&		$1\times10^{-6}$\\
Al$_2$O$_3$ (corundum)&		0.5-3&		$3\times10^{-8}$\\
Si$_3$N$_4$&			$\sim$1&	$2\times10^{-9}$\\
\hline
\end{tabular}
\end{table}
\footnotetext[2]{Mass fraction
in ``primitive'' carbonaceous chondrite
meteorites.}
\footnotetext[3]{SiC and graphite grains sometimes contain very small
TiC, ZrC, and MoC inclusions.}
\setcounter{footnote}{3}

Detectors aboard the Ulysses and Galileo probes measured impacts by
interplanetary dust particles, but also observed impacts 
attributed to the
flux of interstellar grains expected 
due to the 20 km s$^{-1}$ motion of the Sun relative
to the local interstellar medium.
The detectors are sensitive only to collisions with the large-size end
of the interstellar grain distribution, but
the observed 
distribution of impact energies appears to be consistent with both the
overall numbers and size distribution expected based on studies of
interstellar extinction.

The lines of evidence listed above serve to strongly constrain theoretical
models for the interstellar grain population.
Unfortunately, interstellar grain researchers have thus far not been
able to uniquely
determine the grain model from the available observational constraints,
so debates continue concerning the details.

Nevertheless, there is broad consensus on a number of properties of
interstellar grains:
\begin{itemize}
\item The grain population must have a broad size distribution, 
	with grain radii\footnote{%
	The particles are of course not spherical.
	By grain ``radius'' we refer to the radius of a sphere of equal volume.
	}
covering the range 0.5nm -- 300nm -- a factor of
at least $10^8$ in grain mass.
\item Most of the grain mass is in particles with radii
	$50\ltsim a\ltsim300$nm.
\item Most of the grain area is in particles with radii
	$a \ltsim 50$nm.
\item Approximately 50\% of the grain mass in diffuse clouds is contributed by 
	amorphous silicate material, accounting for the broad 9.7$\mu$m and
	18$\mu$m silicate features.
	Most of the silicon, iron, and magnesium abundance is in solid
	form.
\item Approximately 50\% of the grain mass in diffuse clouds is contributed by 
	carbonaceous material.  Approximately 2/3 of the total carbon
	abundance is in solid form.
\item The 217.5nm feature is probably 
	due to some form of carbonaceous material,
	possibly graphitic.
\end{itemize}

\subsection*{Graphite-Silicate Grain Model}

It is not possible to invert the observations to obtain a unique
grain model.
Instead, one makes some assumptions concerning the grain 
composition and
the form of the size distribution, and then attempts to adjust the model
to achieve a good match to the observed interstellar extinction,
infrared emission, and other constraints such as observed gas phase
abundances.

One grain model which has proven fairly successful in conforming to
observations consists of a mixture of carbon grains and
silicate grains.
The carbon grain material is taken to have the optical constants of
crystalline graphite.
A good fit to the extinction can be obtained if both graphite and
silicate grains have size distributions which are approximately
a power-law, $dn/da\propto a^{-3.5}$, truncated at
$a_{\rm min}\approx5\nm$ and $a_{\rm max}\approx 250\nm$
The $217.5\nm$ feature is then reproduced 
by the $a\ltsim15 \nm$ graphite grains.
This grain model -- the combination of graphite and silicate grains,
and the $dn/da\propto a^{-3.5}$ power-law -- was first put forward
in 1977
by Mathis, Rumpl, and Nordsieck, and is often referred to
as the ``MRN'' model.
The MRN model achieves a good fit to the $R_V=3.1$ extinction curve
for diffuse clouds with essentially all of the Mg, Si, and Fe in 
the silicate grains, and approximately 2/3 of the C in the graphite
grains, in reasonable agreement with observed depletions.
The grains are heated to temperatures $\sim$18~K by starlight,
and the resulting thermal emission is approximately consistent with
the observed far-infrared emission at $\lambda > 60\micron$.

\subsection*{Ultrasmall Dust Grains}

Unfortunately, the original graphite-silicate grain model described above
fails to reproduce the $\lambda < 50\micron$ infrared emission
shown in Figure \ref{fig:iremission}.
The model 
lacks the ultrasmall grain component
required to explain the observed $3\ltsim\lambda\ltsim50\micron$ infrared
emission from interstellar dust.
The simplest modification is to allow
the carbon grain distribution to extend down to very 
small sizes, with the smallest grains assumed to have infrared optical
properties suitable to explain the emission features at
3.3, 6.2, 7.6, 8.6, and 11.2$\micron$ when heated to the appropriate
temperatures by absorption of single starlight photons.
In order to have sufficient numbers of ultrasmall grains, while still
reproducing the extinction curve, the size distribution of at least
the carbonaceous grains can no longer be approximated by a single 
power-law.

\subsection*{Other Grain Models}

Other grain models, with different grain geometries and/or compositions,
have been proposed to account for the observed
interstellar extinction and infrared emission.

\begin{itemize}
\item Mathis and Whiffen proposed a grain model wherein the larger grains
are porous aggregates of small graphite and silicate particles.
\item Some authors have favored metal oxides (MgO, SiO, FeO) either in
addition to, or in place of, silicates.
\item Some models assume the silicate grains in diffuse clouds 
to be coated with a carbonaceous ``mantle'' material, which might
be hydrogenated amorphous carbon.
\item Metallic Fe and FeS incorporate significant fractions of the Fe in
some grain models.
\item Small particles of crystalline silicon, with hydrogenated or oxidized
surfaces, have been proposed as an explanation for the observed
luminescence near $\sim$700nm.
\end{itemize}
Grain models generally tend to have 
the bulk of the Fe, Mg, Si in dust, approximately
2/3 of the C, and about 20\% of the O, resulting in dust-to-gas mass
ratios of $\sim$0.007.

\subsection*{Dynamics of Interstellar Grains}

Interstellar grains are acted on by forces due to a number of
distinct physical processes, including:
\begin{itemize}
\item Gas drag forces when the grain velocity differs from that of the gas.
In addition to direct collisions with atoms and ions, there is
also a ``plasma drag'' force on charged grains due to momentum
transfer with ions which do not actually collide with the grain.
For subsonic motion, the plasma drag force is typically a factor 
$\sim$20 -- 30
larger than the drag due to direct collisions with ions.
\item Electromagnetic forces on charged grains.  Since there is usually
a ``plasma'' 
reference frame in which the electric field is very small (interstellar
plasma being a very good conductor), 
the electromagnetic
force can be attributed to the Lorentz force 
${\bf F}=Q({\bf v}/c)\times{\bf B}$ Lorentz force, where
$Q$ is the grain charge, ${\bf B}$ is the magnetic field and
${\bf v}$ is the grain velocity relative to the plasma.
\item Scattering and absorption of photons 
(``radiation pressure'') when the grain is illuminated by an
anisotropic radiation field.
\item Poynting-Robertson drag, as when a grain is moving perpendicular
to a directional radiation field.
While not normally important in the
interstellar medium, Poynting-Robertson drag can be very important for
dust grains orbiting stars (see {\scriptsize INTERPLANETARY DUST}).
\item Recoil forces when photoelectrons or photodesorbed molecules
are emitted anistropically from a grain which is illuminated by
an anisotropic ultraviolet radiation field.
\item Gravitational force.  In clouds supported by gas pressure, dust
grains will tend to ``sediment'' toward the minimum of the gravitational
potential.
\end{itemize}
Because the gas and grains are subject to different forces, the dust grains
generally have a drift velocity relative to the gas.
In typical diffuse clouds, anisotropic starlight can result in drift
velocities of order $\sim$$0.1\kms$.
In regions
with strongly anisotropic ultraviolet radiation fields 
(such as photodissociation regions) the drift velocities can be larger.

The rotational dynamics of interstellar grains are also of great interest,
in connection with both the problem of grain alignment (which requires
alignment of the grain angular momentum vector with the local magnetic
field) and with electric dipole radiation which will be radiated by
very rapidly-rotating ultrasmall grains.
The grain angular momentum is affected by:
\begin{itemize}
\item Collisions with gas atoms and molecules.
\item Recoil associated with photoelectric emission.
\item Recoil associated with H$_2$ formation on the grain surface.
\item Absorption and scattering of starlight by an asymmetric grain.
\item Thermal infrared emission.
\item The interstellar magnetic field acting on the magnetic dipole
	moment resulting from the Barnett effect in a spinning grain.
\item The interstellar magnetic field acting on the magnetization
	induced in the spinning grain by the interstellar magnetic field
	(the ``Davis-Greenstein'' torque associated with paramagnetic
	dissipation).
\end{itemize}
Because the interstellar medium is far from thermodynamic equilibrium, some
of these torques can act systematically.

Grains can be driven to rotational
kinetic energies much larger than $kT_{gas}$ (where $T_{gas}$
is the gas temperature) by collisions with gas atoms and
molecules (because the grain and gas temperatures differ), by
photoelectric emission or by absorption and scattering of starlight 
(because the starlight radiation field is not
in thermodynamic equilibrium with the grain temperature), and by
H$_2$ formation (because the H$_2$ abundance and gas and grain temperatures
are not in thermodynamic equilibrium).

Gradual alignment of the grain angular momentum with the magnetic field
can be produced by paramagnetic dissipation (because the grain rotational
kinetic energy is not in thermodynamic equilibrium with the
``vibrational'' temperature of the grain).

\subsection*{Effects of Interstellar Grains}

\subsubsection*{Photoelectric Heating}

Perhaps the most important effect of interstellar dust grains is their
role in heating the gas via photoelectric emission.
When a dust grain absorbs an ultraviolet photon of energy $h\nu$, there
is a probability $Y(h\nu)$ -- often referred to as the ``yield'' -- 
that a photoelectron will escape from the
grain surface and thermalize with the local gas.
When this happens, the kinetic energy of the emitted electron
(at ``infinity'', since the
grain it is escaping from may be charged)
acts
to heat the interstellar medium.
On average, grains must capture electrons as rapidly as
they are ejected, and the net heating rate then is equal to the
difference between the mean energy of the emitted electrons and the
captured electrons.
Since photoelectric yields $Y$ may be of order $\sim$10\%, and since
the kinetic energy of the electron at infinity may be of order
$\sim 1$eV (if the grain is not highly positively charged)
the photoelectric heating mechanism -- at best -- converts $\sim$few~\% of
the absorbed ultraviolet starlight energy into heating of the gas.
While this $\sim$few~\% conversion efficiency may seem small, the
availability of energy in starlight is so great that photoelectric heating
is generally the dominant heating  mechanism for diffuse interstellar gas.

\subsubsection*{Radiative Forces on Dust Grains}

In astrophysics, anisotropic radiation fields are the rule, rather than the
exception, and these anisotropic radiation fields can have dynamical
consequences for dust and gas.
Circumstellar grains are of course subject to extremely anisotropic
radiation from the star.
At a typical point in the interstellar medium, the nonuniform
distribution of stars in the galaxy, together with patchy obscuration
by interstellar dust, results in starlight with appreciable anisotropy.
The dipole component of the starlight anisotropy may be typically
$\sim$10\%.

When the radiation field is anisotropic, scattering and absorption by the
grain produces a net force on the grain, commonly referred to as the
``radiation pressure'' force.
If the grain is coupled collisionally to the gas by gas drag and perhaps
magnetic fields, then the force exerted on the grains is transmitted to
the gas.
These radiation pressure forces can dominate the dynamics of gas in 
certain regions, such as dust-forming winds from cool stars.

\subsubsection*{H$_2$ Formation on Grains}

Interstellar chemistry largely starts with the catalysis of H$_2$ 
on interstellar grain surfaces
(see {\scriptsize INTERSTELLAR CHEMISTRY}).  
The inferred production rate of H$_2$ in the interstellar medium requires
that an appreciable fraction of the H atoms arriving at interstellar
grains surfaces must leave the grains as part of H$_2$ molecules.
While the details of the kinetics remain uncertain, the overall picture is
broadly as follows: When an H atom in a diffuse cloud
arrives at a grain surface, it has a high probability of
sticking.
The H atom then explores some fraction of the grain surface by
either thermal diffusion or quantum tunnelling, until it either
finds another H atom with which it can recombine to form H$_2$,
or it arrives at a location on the grain surface where the
binding by 
either van der Waals forces (``physisorption'') or formation of a
chemical bond (``chemisorption'') is strong enough that it becomes
trapped.
If such trapping occurs, then the surface coverage of trapped
H atoms builds up until newly-arrived H atoms have an appreciable
probability of reacting with a previously-trapped H atom rather than
becoming trapped themselves, or the adsorbed H atoms are removed
by some other process (e.g., photodesorption or thermal desorption).

\subsubsection*{Coupling Magnetic Fields to Neutral Gas}

Dynamically important magnetic fields are commonly present in interstellar
gas (see {\scriptsize INTERSTELLAR MAGNETIC FIELDS}).
In dense molecular gas with very low fractional ionization,
a large fraction of the ``free'' charge resides on positively and negatively
charged dust grains.
Under these conditions -- which prevail in high density regions in
molecular clouds, and presumably in protostellar disks --
dust grains dominate the coupling of magnetic fields to the gas, since the
neutral gas atoms and molecules themselves are
not directly coupled to the
magnetic field.
The charged dust grains are coupled both to the magnetic field (by
Lorentz forces) and to the neutral gas (by collisional drag).
As a result, the grains will drift with a velocity which is intermediate
between the velocity of the neutral gas and the velocity with which the
magnetic field lines (and plasma) ``slip'' through the neutral gas.
Thus the dust grains determine the rate of ``ambipolar diffusion'' of 
magnetic field lines in dense molecular regions.

Charged dust grains play a similar role in the dynamics of magnetohydrodynamic
shock waves in gas of low fractional ionization but dynamically
significant magnetic fields.

\subsection*{Formation and Destruction of Interstellar Grains}

While interstellar grains are observed to be ubiquitous, it is not
obvious that this should be so.
In the violent interstellar medium
(see {\scriptsize INTERSTELLAR MATTER, SUPERNOVA REMNANTS}),
grains can be destroyed when the gas which they are in is overtaken
by a shock wave with shock speed $v_s\gtsim200\kms$, as is expected
in supernova blastwaves.
Estimates of the rate of occurrence of supernovae, together 
with models for the resulting blastwaves, lead one to estimate that
the mass in interstellar grains will be returned to the gas phase
on a timescale of $\sim 5\times 10^8{\,\rm yr}$, short compared to
the age of the Galaxy and the interstellar medium.
We know that grains are injected into the interstellar medium in
outflows from cool red
giants and supergiants, and even in supernova ejecta, but
if there were no conversion of gaseous atoms
back to solid form in the interstellar medium, we would expect very little
grain material to be present at any time.
In particular, even for elements like Fe we would expect most of the
atoms to be in the gas phase.
The fact that this is not so -- that most of the interstellar Fe is in
fact missing from the gas phase -- requires that there be efficient
recondensation of Fe, Si, Ca, and other elements back into solid form
{\it in the interstellar medium}.
The ``mineralogy'' of interstellar grains must therefore largely reflect
the surface chemistry which will occur on the surfaces of interstellar
grains.

\subsection*{Bibliography}

{\it Dust in the Galactic Environment}, by D.C.B. Whittet
(London: IOP Publishing; 1992), provides an
excellent overall description of interstellar dust grains.

The observed optical properties of dust have been recently reviewed
by J.S. Mathis (1990), {\it Ann. Rev. Astr. Astrophys.}, {\bf28}, 37.

The ultrasmall grain population is discussed by Puget, J.L., \&
Leger, A. (1989), {\it Ann. Rev. Astr. Astrophys.}, {\bf 27}, 161.

An overall review of interstellar and circumstellar dust is given by
Dorschner, J., \& Henning, T. (1995), {\it Astr. Astrophys. Rev.},
{\bf 6}, 271.

Microwave emission from dust grains is reviewed by
Draine, B.T., \& Lazarian, A. (1999), in
{\it Microwave Foregrounds}, ed. A. de Oliveira-Costa \& M. Tegmark
(San Francisco: Astron. Soc. Pacific), p. 133

\bigskip
\noindent BRUCE T. DRAINE
\end{document}